\documentclass[aps,prb,twocolumn,superscriptaddress]{revtex4}
\usepackage[active]{srcltx}
\usepackage{graphicx}
\usepackage{amsmath,amssymb}
\begin{document}

%\preprint{}

\title{Origin of the ESR spectrum in the Prussian Blue analogue RbMn[Fe(CN)$_{6}$]$\cdot$ H$_{2}$O }


\author{\'A. Antal}
\email{agnesantal@gmail.com}
\affiliation{Budapest University of Technology and Economics, Institute of Physics and Condensed Matter Research Group of the Hungarian Academy of Sciences, P.O. Box 91, H-1521 Budapest, Hungary}

\author{A. J\'anossy}

%\thanks{}
\affiliation{Budapest University of Technology and Economics, Institute of Physics and Condensed Matter Research Group of the Hungarian Academy of Sciences, P.O. Box 91, H-1521 Budapest, Hungary}

\author{L. Forr\'o}

\affiliation{Institute of Physics of Complex Matter, FBS, Swiss Federal Institute of Technology (EPFL), CH-1015 Lausanne, Switzerland}

\author{E. J. M. Vertelman}
\affiliation{Zernike Institute for Advanced Materials, University of Groningen, Nijenborgh 4, 9747 AG Groningen, The Netherlands}

\affiliation{Stratingh Institute of Chemistry, University of Groningen, Nijenborgh 4, 9747 AG Groningen, The Netherlands}

\author{P. J. van Koningsburgen}

\affiliation{Chemical Engineering and Applied Chemistry School of Engineering and Applied Science, Aston University, Birmingham B4 7ET, England}

\author{P.H.M. van Loosdrecht}

\affiliation{Zernike Institute for Advanced Materials, University of Groningen, Nijenborgh 4, 9747 AG Groningen, The Netherlands}


\date{\today}

\begin{abstract}

We present an ESR study at excitation
frequencies of 9.4~GHz and 222.4~GHz of powders and single crystals of a Prussian Blue analogue (PBA), RbMn[Fe(CN)$_{6}$]$\cdot$ H$_{2}$O in which Fe and Mn undergoes a charge transfer transition between 175 and 300 K.  The ESR of PBA powders, also reported by Pregelj et al.\cite{Pregelj2007} is assigned to
cubic magnetic clusters of Mn$^{2+}$ ions surrounding Fe(CN)$_6$ vacancies. The clusters are well isolated from the bulk and are superparamagnetic below 50 K. In single crystals various defects with lower symmetry are also observed. Spin-lattice relaxation broadens the bulk ESR beyond observability. This strong spin relaxation is unexpected above the charge transfer transition and is attributed to a mixing of the Mn$^{3+}$-Fe$^{2+}$ state into the prevalent Mn$^{2+}$-Fe$^{3+}$ state.
\end{abstract}

%\keywords{}

\maketitle

%\section{}
%\subsection{}
%\subsubsection{}

\section{Introduction}

The interest in the extensively studied Prussian Blue analogues\cite{Vertelman2008, Moritomo2002, Lummen2008}, A(I)M(II)[N(III)(CN)$_6$], where A=(Na, K, Rb, Cs); M=(Mn, Co, Cr); N= (Fe, Cr), lies in a charge transfer transition between high temperature (HT) and low temperature (LT) phases \cite{Moritomo2002,Tokoro2008, Kato2003}. Besides a temperature change, the transition can be triggered by illumination with visible light \cite{Tokoro2003, Moritomo2003a}, or X-rays \cite{Margadonna2004} and also by application of pressure \cite{Moritomo2003a}. The materials have potential technological applications like magneto-optic devices based on photoinduced magnetism.

In the compound studied in this paper, \mbox{RbMn[Fe(CN)$_{6}$]$\cdot$H$_{2}$O}, the HT phase is cubic (F$\bar{4}$3m; Z=4) with Mn$^{2+}$ and Fe$^{3+}$ ions in S=5/2 and S=1/2 spin states respectively. The LT Mn$^{3+}$ (S=2), Fe$^{2+}$ (S=0) phase has a tetragonal symmetry (I$\bar{4}$m2; Z=2).
The charge transfer transformation is driven by a Jahn-Teller distortion of Mn$^{3+}$ ions \cite{Moritomo2002}.
The transition has a broad thermal hysteresis with the LT$\rightarrow$HT transition at approximately 300 K and the HT$\rightarrow$LT transition
at about 175 K \cite{Vertelman2008}.
In the LT phase, below $T_F=11 K$ the S=2 spins of the Mn$^{3+}$ ions order ferromagnetically
 \cite{Moritomo2003, Ohkoshi2007}.

\begin{figure}
\includegraphics[angle=0,width=8.0 cm]{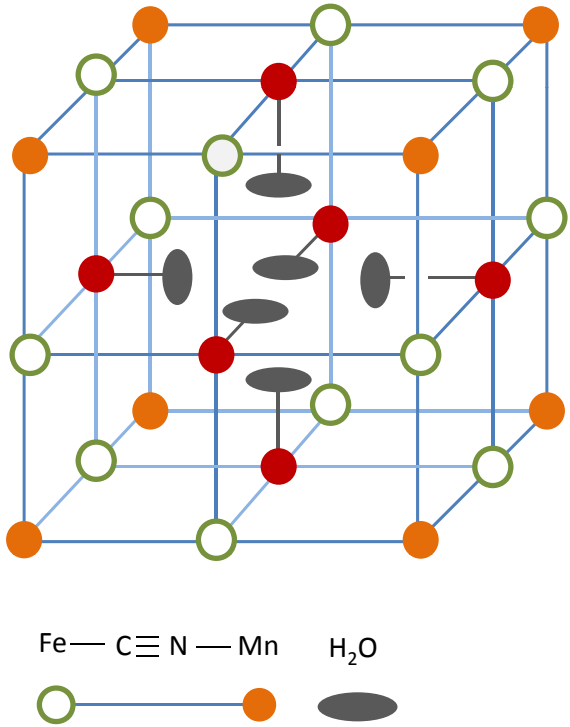}
\caption{Structure of a cubic Mn$^{2+}$ cluster in RbMn[Fe(CN)$_{6}$]$\cdot$ H$_{2}$O surrounding an Fe(CN)$_6$ vacancy. Rb$^{+}$ ions and
noncoordinated lattice water molecules are omitted for clarity \cite{Vertelman2006}. The 6 Mn$^{2+}$ ions of the cluster are drawn darker. \label{fig4}}
\end{figure}

In the present detailed study of single crystals and powder PBA samples, we discuss the origin of ESR active defects and set a lower limit for the unusually fast spin relaxation rate in the bulk. The experimental findings in powders are in good agreement with the study of Pregelj et al.\cite{Pregelj2007}. We assign an isotropic ESR line to a Cubic cluster (denoted hereafter "C-cluster") of Mn$^{2+}$ ions surrounding the structural defect shown in FIG. \ref{fig4} and described earlier by Vertelman et al.\cite{Vertelman2008}. The C-clusters are remarkably well isolated from the bulk and are only little affected by the HT-LT charge transfer transition. At low temperatures C-clusters remain superparamagnetic even below the ferromagnetic Curie temperature of the bulk, at $T_F$=11 K. In single crystals, defects with lower symmetry are also observed.
We find no ESR of the bulk in spite of an intensive search; spin relaxation broadening of the line is more than 1 T.
The lack of a bulk ESR in the high temperature phase is surprising as the S=5/2 spin Mn$^{2+}$ and the S=1/2 Fe$^{3+}$ ions are usually easily observed. We argue that an admixture of the Mn$^{3+}$-Fe$^{2+}$ state into the Mn$^{2+}$-Fe$^{3+}$ state is the reason for the fast spin relaxation in the HT phase.

\section{Experimental results}

The RbMn[Fe(CN)$_6$]$\cdot$H$_2$O single crystals are from the same batch as in reference \cite{Vertelman2008}; details of the synthesis and characterization are given in the Appendix.
Measurements at 222.4 GHz were performed by a home built ESR spectrometer while at 9.4 GHz we used a Bruker
ELEXSYS E500 spectrometer.
The sample was annealed at 320~K before thermal cycling experiments to ensure that it was fully in the high spin state. In powdered samples a single strong ESR line is observed at both 9.4 and 222.4 GHz and at all temperatures between 4~K and 320~K.

\begin{figure}
\includegraphics[angle=0,width=9.3 cm]{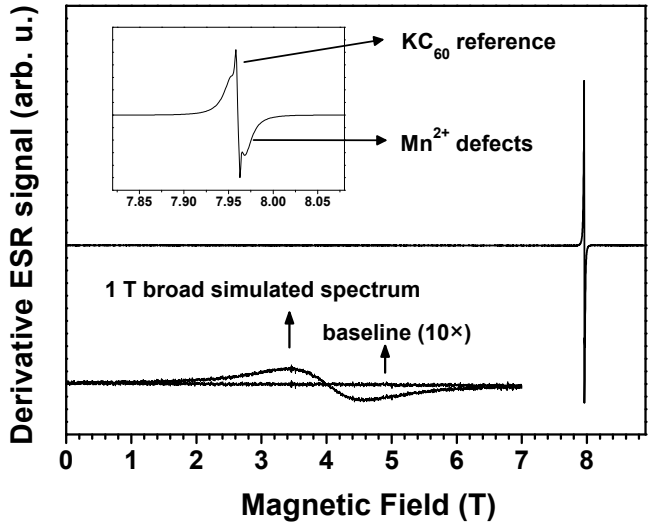}
\caption{ 222.4~GHz ESR spectra of PBA powder at 175~K. Inset: the same spectrum zoomed around g=2. The flat, 10
times magnified baseline and the superimposed simulated spectra show that a 1~T broad ESR with the expected intensity would be easily detected in the 0-9 T field range.\label{fig1}}
\end{figure}

%222.4GHz, Hm=2mT modulation GHz of a PB powder sample (EV 0707).  RbMnFe(CN)6.H2o single crystal

\begin{figure}
\includegraphics[angle=0,width=9.1cm]{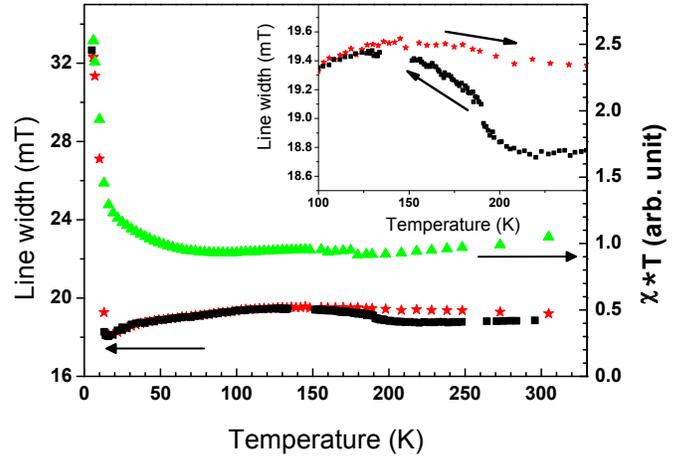}
\caption{  9.4~GHz ESR line width (stars: heating, cubes: cooling) and  ESR Susceptibility times temperature  (triangles) of PBA powder sample
 as a function of temperature. Inset: hysteresis of the line width at the HT$\rightarrow$LT transition. \label{fig3}}
\end{figure}
% EV0707

 The intensity of the ESR line, which is proportional to the spin susceptibility of the ESR active species, was measured at ambient temperature in a powder sample
 at 9.4~GHz using CuSO$_{4}$$ \cdot $5H$_{2}$O  reference. The ESR intensity of the powder sample corresponds
 to a concentration of about 2.5\% of Mn$^{2+}$ ions. This is much less than
 expected for a resonance arising from one S=5/2 Mn$^{2+}$ paramagnetic ion per formula unit. A line with so small intensity cannot arise from the ESR of the bulk, only from defect sites.

 We searched in vain for a broad line corresponding to
 the bulk of magnetic ions. As shown in FIG.~\ref{fig1}, at 222.4~GHz the 20~mT broad resonance line of Mn$^{2+}$ defect sites at g=2.0006 is the only observable resonance within the range of 0 to
 9~T.
 Thus the ESR of the bulk is very broad: a 1~T broad resonance with the intensity of the bulk Mn$^{2+}$ ions would have been easily detected as shown in FIG.~\ref{fig1}.

    The defects with isotropic ESR were observed in powders at both frequencies and all temperatures. At room temperature, the line width of the defects, $\Delta H$, in the powder is almost independent of the Larmor frequency, $\nu_{L}$.
 On the other hand, the g factor of the powder has a curious frequency dependence, it varies from g=2.022 at
 $\nu_{L}$ =9.4~GHz to 2.0006 at 222.4~GHz. The HT-LT transition affects only weakly the ESR signal of the defect sites. The line is slightly narrower
in the HT phase than in the LT phase, and there is a small hysteresis in the line width between 175 and 300~K (FIG.~\ref{fig3} inset). Within experimental accuracy the ESR intensity follows a Curie
law (FIG.~\ref{fig3}) between 50 and 300~K, it is not affected by the HT-LT transition. On the other hand, the spin crossover is clear in static magnetization measurements of similar powders. We conclude
that there is no spin crossover in the magnetic ions of the ESR active defect sites.

The defect is not a single isolated magnetic ion. Below 50~K the ESR magnetic spin susceptibility increases faster than predicted by the Curie-law and at 5~K it is several times larger than expected for a constant concentration of free spins.

The 9.4~GHz ESR line slightly narrows
below 50~K and then increases rapidly as the sample is cooled below the ferromagnetic
ordering temperature of the bulk at 11~K (FIG.~\ref{fig3}). At 222.4~GHz the line starts to broaden already at 25~K, i.e at a much higher
temperature than in low magnetic field.

\begin{figure}
\includegraphics[angle=0,width=9.2cm]{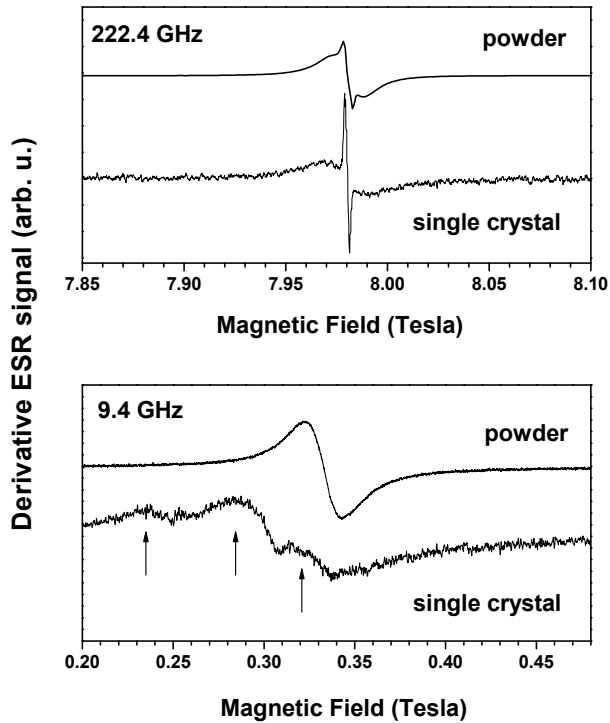}
\caption{   ESR spectra of a PBA single crystal and powder at 222.4~GHz (top) and at 9.4~GHz (bottom) at ambient temperature. For the single crystal spectra the magnetic field is along a principle crystallographic axis. The narrow lines in the 222.4~GHz spectra are from overmodulated KC$_{60}$ reference. The weak 9.4 GHz single crystal ESR lines, indicated by arrows, are superimposed on a broad instrumental background.  \label{fig2}}
\end{figure}

 As in powders, there is no intense resonance of the bulk in the superior sample quality single crystals.
However, unlike in powders, in single crystals two types of ESR active defects appear with isotropic and anisotropic g factors. Three resonances were clearly resolved with
different line widths and resonance fields in the single crystal, 9.4~GHz, ambient temperature ESR spectra (FIG.~\ref{fig2}). The anisotropic defects have a complicated behavior, which is not understood. The line positions and widths are strongly temperature dependent. At temperatures below 110 K, the most intensive line is at about g=2.079 (not shown). The low symmetry defects in the single crystal were only observed at 9.4 GHz.
At 222.4~GHz in a single crystal only an isotropic line was observed at g=2.0006 with a line width similar to the line of the powder spectra.

%EV0712

\section{Discussion}

\subsection{ESR active defects}

In this and the following sections we focus on the properties of defect sites, the reason for the absence of the ESR of the bulk is discussed in subsection \ref{absence}. The g-factor of the ESR in powders is isotropic, otherwise the line would broaden at high frequencies. In the same defects the crystal field (zero field splitting) is small, otherwise the low frequency line would broaden. It is natural to assign this isotropic line to a defect with cubic symmetry. The anisotropic ESR lines in single crystals at low frequency arise from defects with lower symmetry.

We propose that the isotropic g-factor line is the ESR of C-clusters (FIG. \ref{fig4}) with a simple configuration of 6 weakly interacting Mn$^{2+}$ ions on the cube surrounding a defect. A larger cubic cluster with many more ions cannot be entirely ruled out. H$_{2}$O molecules attached to Mn ions fill
the Fe(CN)$_{6}$ vacancy and in this environment the Mn$^{2+}$ state is stabilized at all temperatures.
The clusters have very little magnetic interaction with the surrounding lattice and the ESR is almost unaffected by the HT-LT phase transition. We do not know how this isolation is realized.
The structure of the C-cluster explains qualitatively several characteristics
of the ESR spectra: i.) the isotropy of the g factor, ii.) the nonlinear variation of the line position with frequency, iii) the temperature and frequency dependence of the line width, iv.) the temperature dependence of the magnetic susceptibility of the ESR active sites.

We first discuss i.) and ii.) which follow from the particular structure of the defect, while iii.) and iv.) discussed in subsection \ref{superpara}.
are linked to the superparamagnetism below 50~K.

The C-cluster as a whole is cubic, but individual Mn$^{2+}$ ions are not in a cubic environment since one of their six first Fe neighbors is missing. If Mn$^{2+}$ ions in the C-cluster were isolated from each other then the crystal field (fine structure) and g factor anisotropy would render their ESR lines strongly anisotropic. A magnetic field in a general orientation would split the ESR of C-clusters of a single crystal into 3 Mn$^{2+}$ lines with different g-factors and a fine structure (and a hyperfine structure) of several lines. In the powder, the ESR of noninteracting Mn ions would be broad and frequency dependent. A coupling between Mn$^{2+}$ ions within the C-clusters explains the observed isotropy. An isotropic exchange interaction between Mn ions, mediated by Fe, narrows the g factor anisotropy, $\Delta g/g\cdot \nu_{L}$, and the fine structure splitting, $\nu_{D}$. A small exchange energy, $J$, is sufficient to merge all lines of a single cluster into a single isotropic line. $J>\Delta g/g\cdot h \nu_{L}$ and $J>h \nu_{D}$,
are the conditions for this type of exchange narrowing \cite{Anderson1954}, %Anderson%
At high frequencies (compared to $\nu_{D}$) and high temperatures ($k_B T > h \nu_{L} $) the crystal field (or "zero field splitting") does not shift the line.
On the other hand, at low ESR frequencies, where $\nu_{L}$ and $\nu_{D}$ are comparable, the crystal field shifts the average line position even if the spectrum is narrowed by exchange. In second order, the fine structure shift is of the order of $\nu_{D}^{2}/\nu_{L}$. The observed apparent g-shift from 2.0006 at 222.4 GHz to 2.022 at 9.4 GHz corresponds to a fine structure splitting of $|\nu_{D}|/ \gamma=0.07$~T if the shift is from the -1/2$\rightarrow$1/2 transition of Mn$^{2+}$ ions. (The shifts of other transitions are of similar order.)

The crystal field at cluster Mn sites changes the 222.4~GHz spectra at low temperatures where $k_B T < h \nu_{L} $. At high temperatures the thermal population of Zeeman levels are not very different, thus ESR fine structure transitions have comparable intensities. However, at low temperatures the thermal energy (k$_B$T) is smaller than the Zeeman energy and the higher Zeeman energy states are depopulated. In this case only fine structure transitions between the lower Zeeman energy states contribute to the spectrum. As a result the exchange narrowed line shifts towards the fine structure transition between the lowest lying states.\cite{Nagy2009}. We have observed this shift at 222.4 GHz where $5h\nu_{L}$=40 K is the energy of the highest Zeeman level. The resonance field of the PB powder at 25 K is down shifted 0.03~T compared to ambient temperature. As expected, this shift at 25K and 222.4 GHz is of the order of the estimation of $|\nu_{D}|/ \gamma=0.07$~T from the low frequency data. From the sign of the shift we conclude that $\nu_{D}<0$.
The complicated structure of single crystal 9.4~GHz spectra at room temperature can be caused by the crystal field and g-factor anisotropy in non cubic clusters.

 As explained in subsection \ref{absence},
spin relaxation of bulk magnetic ions in the HT state is extremely fast and is without doubt even faster in the LT state. Nevertheless, the HT to LT transition changes the line width by only 2~mT (FIG. \ref{fig3}), thus C-clusters are well isolated from the bulk. A significant coupling to the bulk would result in a fast spin relaxation and a large ESR line broadening.
As expected for isolated C-clusters with no phase transition; the ESR intensity is also unchanged at the HT$\rightarrow$LT transition.

 \subsection{Superparamagnetism} \label{superpara}

Above 50~K the spin susceptibility of defects is to a good approximation Curie like, but below 50 K the ESR intensity increases faster than 1/T.
We argue that this is due superparamagnetism at low temperatures where a ferromagnetic exchange between Mn ions in clusters is significant. For definitiveness, we consider the cluster of 6 S=5/2 Mn$^{2+}$ ions coupled through 12 S=1/2 Fe$^{3+}$ ions (FIG. \ref{fig4}), other configurations with a larger cluster size cannot be ruled out. The cluster has a common exchange narrowed resonance of all magnetic ions since the g-factor is about 2 for both Mn$^{2+}$ and Fe$^{3+}$ ions.
An indirect Mn - Mn exchange coupling through magnetic Fe$^{3+}$ ions is always ferromagnetic, independent of the sign of the Mn-Fe exchange. At low temperatures the total spin of six Mn$^{2+}$ ions is S=15 and the spin of the full cluster is between S$_{C}$=21, and S$_{C}$=9 for ferromagnetic and antiferromagnetic Mn-Fe coupling respectively. Thus at low temperatures, the magnetic moment of the cluster is large and the susceptibility increases with decreasing temperature much faster than for non-interacting ions. At much higher temperatures than J the susceptibility is about that of free Mn$^{2+}$ ions. A Mn-Mn exchange interaction within the cluster of the order of J=10~K (and a much weaker interaction with the bulk) explains the ESR susceptibility.

The ferromagnetic transition of the bulk at $T_F$=11~K does not affect significantly the 9 GHz ESR intensity, which continuously increases to the lowest measurement temperature of 5 K. On the other hand, below 11~K the 9 GHz ESR line width increases abruptly. This is well explained with a ferromagnetic transition of the bulk at 11~K and no ferromagnetic ordering of the superparamagnetic clusters. The cluster ESR line broadening of 14~mT below the ferromagnetic transition of the bulk arises either from long range dipolar interactions, i.e. inhomogeneous demagnetizing fields or a small exchange coupling to the bulk.

\subsection{Absence of ESR of the bulk} \label{absence}

 We discuss the possible reasons for which only defects are ESR active and the ESR of the bulk material was not observed.

In general, to observe ESR the spin relaxation rate must be less than $\nu_{L}$.
The lack of a spin resonance in the LT phase is not surprising:
  Fe$^{2+}$ is not magnetic and Mn$^{3+}$ has an S=2 spin for which orbital effects are important.
 Phonons modulate the crystal field and the fast spin relaxation broadens the ESR of Mn$^{3+}$ ions beyond observability. On the other hand, the lack of bulk ESR above the spin crossover is not easily explained.
 Crystals with Mn$^{2+}$ (S=5/2) and  Fe$^{3+}$ (S=1/2) ions usually have narrow ESR lines with giromagnetic
 factors near g=2. Fine splitting from crystal fields is relatively small for the
 half filled 3d$^{5}$ shell of Mn$^{2+}$. There is no zero field splitting for
 S=1/2 Fe$^{3+}$ ions either, the ESR of this ion is not strongly anisotropic and has been frequently observed in solids. Moreover, crystal
 field anisotropy (fine structure splitting) and the dipolar interaction are ineffective in magnetically dense systems.
 In PBA the exchange interaction between Mn and Fe ions is larger than dipolar and single ion crystal
 field energies and the ESR is exchange narrowed i.e. one expects a common Mn$^{2+}$, Fe$^{3+}$ narrow ESR resonance in the bulk.
 Yet, we find that the ESR
 of the bulk of PBA is very broad, more than 1~T, i.e. at least two orders
 of magnitude broader than the ESR of the Mn$^{2+}$ defects clusters in the same system. At a Larmor frequency of 9.4~GHz, a larger than 1~T line width means that
 the life time broadening is more than the resonance field and there is no ESR of the bulk at all.

 A small admixture of the S=2 Mn$^{3+}$ and S=0 Fe$^{2+}$ states into the pure
 S=5/2 Mn$^{2+}$ and S=1/2 Fe$^{3+}$ states in the HT phase is the most
 probable explanation for this rapid spin lattice relaxation. Charge fluctuations persisting above the LT - HT transition induce a rapid spin
 relaxation and consequently a very broad or non-existent ESR.
 This admixture was also seen by X-ray Photoemission Spectroscopy
 (XPS) and Raman spectroscopy \cite{Lummen2008}.

\section{Conclusion}

 Single crystals of \mbox{RbMn[Fe(CN)$_{6}$]$\cdot$H$_{2}$O} contain ESR active defects both with cubic and lower symmetry. At high frequency and in powders the low symmetry defects are not observed. The isotropic defect ESR is assigned to a cubic C-cluster of Mn$^{2+}$ ions around Fe(CN)$_{6}$ vacancies, which interacts very weakly with the bulk. At low temperatures the C-clusters are superparamagnetic but they do not order at
 the $T_F$=11~K ferromagnetic transition of the bulk. We suggest that the electronic configuration at high temperatures is not a pure configuration of  Mn$^{2+}$ and Fe$^{3+}$ ions but has an
admixture of Mn$^{3+}$ and Fe$^{2+}$ states of the low temperature phase. This is the reason for the lack of an observable ESR of Mn$^{2+}$ and Fe$^{3+}$ ions in the bulk.

\appendix*
\section{}

The composition of the powdered sample was \mbox{(Rb$_{0.91}$Mn[Fe(CN)$_6$]$_{0.97}$$\cdot$ 1.53H$_2$O)}. A solution of 0.495~g MnCl$_2$$\cdot$4H$_2$O in 25~mL of H$_2$O was added to a mixed solution of 0.823~g K$_3$[Fe(CN)$_6$] and 3.023~g RbCl in 25~mL of H$_2$O. The addition speed was kept constant at 6~mL$\cdot$h$^{-1}$ with a syringe pump (Sage Instruments, model 352). The solution was stirred at a constant rate of 5.5~rps. The temperature of the combined solution was kept constant at 318~K with a water bath. A brown powder precipitated which was centrifuged and washed twice with distilled water of room temperature. The samples were dried overnight in vacuum. Elemental analysis was performed at the analysis facility of CNRS in Vernaison, France. Calculated for Rb$_{0.91}$Mn[Fe(CN)$_{6}$]$_{0.97}$$\cdot$1.53H$_2$O: Rb 21.26$\%$, Mn 15.03$\%$, Fe 14.80$\%$, C 19.10$\%$, N 22.27$\%$, H 0.84$\%$. Found: Rb 21.26$\%$, Mn 15.03$\%$, Fe 14.80$\%$, C 20.23$\%$, N 22.63$\%$, H 0.32$\%$. Yield (based on Mn): 84$\%$.

\begin{acknowledgments}
This work was supported by the Hungarian National Research Fund OTKA K68807.
The authors thank Prof. Palstra for the use of the X-ray powder diffractometer and the MPMS.
The work in Lausanne was supported by the Swiss NSF and its NCCR MaNEP.
\end{acknowledgments}


\begin{thebibliography}{14}
\expandafter\ifx\csname natexlab\endcsname\relax\def\natexlab#1{#1}\fi
\expandafter\ifx\csname bibnamefont\endcsname\relax
  \def\bibnamefont#1{#1}\fi
\expandafter\ifx\csname bibfnamefont\endcsname\relax
  \def\bibfnamefont#1{#1}\fi
\expandafter\ifx\csname citenamefont\endcsname\relax
  \def\citenamefont#1{#1}\fi
\expandafter\ifx\csname url\endcsname\relax
  \def\url#1{\texttt{#1}}\fi
\expandafter\ifx\csname urlprefix\endcsname\relax\def\urlprefix{URL }\fi
\providecommand{\bibinfo}[2]{#2}
\providecommand{\eprint}[2][]{\url{#2}}

\bibitem[{\citenamefont{Pregelj et~al.}(2007)\citenamefont{Pregelj, Zorko,
  Arcon, Margadonna, Prassides, van Tol, Brunel, and Ozarowski}}]{Pregelj2007}
\bibinfo{author}{\bibfnamefont{M.}~\bibnamefont{Pregelj}},
  \bibinfo{author}{\bibfnamefont{A.}~\bibnamefont{Zorko}},
  \bibinfo{author}{\bibfnamefont{D.}~\bibnamefont{Arcon}},
  \bibinfo{author}{\bibfnamefont{S.}~\bibnamefont{Margadonna}},
  \bibinfo{author}{\bibfnamefont{K.}~\bibnamefont{Prassides}},
  \bibinfo{author}{\bibfnamefont{H.}~\bibnamefont{van Tol}},
  \bibinfo{author}{\bibfnamefont{L.~C.} \bibnamefont{Brunel}},
  \bibnamefont{and}
  \bibinfo{author}{\bibfnamefont{A.}~\bibnamefont{Ozarowski}},
  \bibinfo{journal}{Journal Of Magnetism And Magnetic Materials}
  \textbf{\bibinfo{volume}{316}}, \bibinfo{pages}{E680} (\bibinfo{year}{2007}).

\bibitem[{\citenamefont{Vertelman et~al.}(2008)\citenamefont{Vertelman, Lummen,
  Meetsma, Bouwkamp, Molnar, van Loosdrecht, and van
  Koningsbruggen}}]{Vertelman2008}
\bibinfo{author}{\bibfnamefont{E.~J.~M.} \bibnamefont{Vertelman}},
  \bibinfo{author}{\bibfnamefont{T.~T.~A.} \bibnamefont{Lummen}},
  \bibinfo{author}{\bibfnamefont{A.}~\bibnamefont{Meetsma}},
  \bibinfo{author}{\bibfnamefont{M.~W.} \bibnamefont{Bouwkamp}},
  \bibinfo{author}{\bibfnamefont{G.}~\bibnamefont{Molnar}},
  \bibinfo{author}{\bibfnamefont{P.~H.~M.} \bibnamefont{van Loosdrecht}},
  \bibnamefont{and} \bibinfo{author}{\bibfnamefont{P.~J.} \bibnamefont{van
  Koningsbruggen}}, \bibinfo{journal}{Chemistry Of Materials}
  \textbf{\bibinfo{volume}{20}}, \bibinfo{pages}{1236} (\bibinfo{year}{2008}).

\bibitem[{\citenamefont{Moritomo et~al.}(2002)\citenamefont{Moritomo, Kato,
  Kuriki, Takata, Sakata, Tokoro, Ohkoshi, and Hashimoto}}]{Moritomo2002}
\bibinfo{author}{\bibfnamefont{Y.}~\bibnamefont{Moritomo}},
  \bibinfo{author}{\bibfnamefont{K.}~\bibnamefont{Kato}},
  \bibinfo{author}{\bibfnamefont{A.}~\bibnamefont{Kuriki}},
  \bibinfo{author}{\bibfnamefont{M.}~\bibnamefont{Takata}},
  \bibinfo{author}{\bibfnamefont{M.}~\bibnamefont{Sakata}},
  \bibinfo{author}{\bibfnamefont{H.}~\bibnamefont{Tokoro}},
  \bibinfo{author}{\bibfnamefont{S.}~\bibnamefont{Ohkoshi}}, \bibnamefont{and}
  \bibinfo{author}{\bibfnamefont{K.}~\bibnamefont{Hashimoto}},
  \bibinfo{journal}{Journal Of The Physical Society Of Japan}
  \textbf{\bibinfo{volume}{71}}, \bibinfo{pages}{2078} (\bibinfo{year}{2002}).

\bibitem[{\citenamefont{Lummen et~al.}(2008)\citenamefont{Lummen, Gengler,
  Rudolf, Lusitani, Vertelman, van Koningsbruggen, Knupfer, Molodtsova,
  Pireaux, and van Loosdrecht}}]{Lummen2008}
\bibinfo{author}{\bibfnamefont{T.~T.~A.} \bibnamefont{Lummen}},
  \bibinfo{author}{\bibfnamefont{R.~Y.~N.} \bibnamefont{Gengler}},
  \bibinfo{author}{\bibfnamefont{P.}~\bibnamefont{Rudolf}},
  \bibinfo{author}{\bibfnamefont{F.}~\bibnamefont{Lusitani}},
  \bibinfo{author}{\bibfnamefont{E.~J.~M.} \bibnamefont{Vertelman}},
  \bibinfo{author}{\bibfnamefont{P.~J.} \bibnamefont{van Koningsbruggen}},
  \bibinfo{author}{\bibfnamefont{M.}~\bibnamefont{Knupfer}},
  \bibinfo{author}{\bibfnamefont{O.}~\bibnamefont{Molodtsova}},
  \bibinfo{author}{\bibfnamefont{J.-J.} \bibnamefont{Pireaux}},
  \bibnamefont{and} \bibinfo{author}{\bibfnamefont{P.~H.~M.} \bibnamefont{van
  Loosdrecht}}, \bibinfo{journal}{The Journal of Physical Chemistry C}
  \textbf{\bibinfo{volume}{112}}, \bibinfo{pages}{14158}
  (\bibinfo{year}{2008}).

\bibitem[{\citenamefont{Tokoro et~al.}(2008)\citenamefont{Tokoro, Matsuda,
  Nuida, Moritomo, Ohoyama, Dangui, Boukheddaden, and Ohkoshi}}]{Tokoro2008}
\bibinfo{author}{\bibfnamefont{H.}~\bibnamefont{Tokoro}},
  \bibinfo{author}{\bibfnamefont{T.}~\bibnamefont{Matsuda}},
  \bibinfo{author}{\bibfnamefont{T.}~\bibnamefont{Nuida}},
  \bibinfo{author}{\bibfnamefont{Y.}~\bibnamefont{Moritomo}},
  \bibinfo{author}{\bibfnamefont{K.}~\bibnamefont{Ohoyama}},
  \bibinfo{author}{\bibfnamefont{E.~D.~L.} \bibnamefont{Dangui}},
  \bibinfo{author}{\bibfnamefont{K.}~\bibnamefont{Boukheddaden}},
  \bibnamefont{and} \bibinfo{author}{\bibfnamefont{S.-i.}
  \bibnamefont{Ohkoshi}}, \bibinfo{journal}{Chemistry of Materials}
  \textbf{\bibinfo{volume}{20}}, \bibinfo{pages}{423} (\bibinfo{year}{2008}).

\bibitem[{\citenamefont{Kato et~al.}(2003)\citenamefont{Kato, Moritomo, Takata,
  Sakata, Umekawa, Hamada, Ohkoshi, Tokoro, and Hashimoto}}]{Kato2003}
\bibinfo{author}{\bibfnamefont{K.}~\bibnamefont{Kato}},
  \bibinfo{author}{\bibfnamefont{Y.}~\bibnamefont{Moritomo}},
  \bibinfo{author}{\bibfnamefont{M.}~\bibnamefont{Takata}},
  \bibinfo{author}{\bibfnamefont{M.}~\bibnamefont{Sakata}},
  \bibinfo{author}{\bibfnamefont{M.}~\bibnamefont{Umekawa}},
  \bibinfo{author}{\bibfnamefont{N.}~\bibnamefont{Hamada}},
  \bibinfo{author}{\bibfnamefont{S.}~\bibnamefont{Ohkoshi}},
  \bibinfo{author}{\bibfnamefont{H.}~\bibnamefont{Tokoro}}, \bibnamefont{and}
  \bibinfo{author}{\bibfnamefont{K.}~\bibnamefont{Hashimoto}},
  \bibinfo{journal}{Physical Review Letters} \textbf{\bibinfo{volume}{91}},
  \bibinfo{pages}{255502} (\bibinfo{year}{2003}).

\bibitem[{\citenamefont{Tokoro et~al.}(2003)\citenamefont{Tokoro, ichi Ohkoshi,
  and Hashimoto}}]{Tokoro2003}
\bibinfo{author}{\bibfnamefont{H.}~\bibnamefont{Tokoro}},
  \bibinfo{author}{\bibfnamefont{S.}~\bibnamefont{ichi Ohkoshi}},
  \bibnamefont{and}
  \bibinfo{author}{\bibfnamefont{K.}~\bibnamefont{Hashimoto}},
  \bibinfo{journal}{Applied Physics Letters} \textbf{\bibinfo{volume}{82}},
  \bibinfo{pages}{1245} (\bibinfo{year}{2003}).

\bibitem[{\citenamefont{Moritomo
  et~al.}(2003{\natexlab{a}})\citenamefont{Moritomo, Hanawa, Ohishi, Kato,
  Takata, Kuriki, Nishibori, Sakata, Ohkoshi, Tokoro et~al.}}]{Moritomo2003a}
\bibinfo{author}{\bibfnamefont{Y.}~\bibnamefont{Moritomo}},
  \bibinfo{author}{\bibfnamefont{M.}~\bibnamefont{Hanawa}},
  \bibinfo{author}{\bibfnamefont{Y.}~\bibnamefont{Ohishi}},
  \bibinfo{author}{\bibfnamefont{K.}~\bibnamefont{Kato}},
  \bibinfo{author}{\bibfnamefont{M.}~\bibnamefont{Takata}},
  \bibinfo{author}{\bibfnamefont{A.}~\bibnamefont{Kuriki}},
  \bibinfo{author}{\bibfnamefont{E.}~\bibnamefont{Nishibori}},
  \bibinfo{author}{\bibfnamefont{M.}~\bibnamefont{Sakata}},
  \bibinfo{author}{\bibfnamefont{S.}~\bibnamefont{Ohkoshi}},
  \bibinfo{author}{\bibfnamefont{H.}~\bibnamefont{Tokoro}},
  \bibnamefont{et~al.}, \bibinfo{journal}{Phys. Rev. B}
  \textbf{\bibinfo{volume}{68}}, \bibinfo{pages}{144106}
  (\bibinfo{year}{2003}{\natexlab{a}}).

\bibitem[{\citenamefont{Margadonna et~al.}(2004)\citenamefont{Margadonna,
  Prassides, and Fitch}}]{Margadonna2004}
\bibinfo{author}{\bibfnamefont{S.}~\bibnamefont{Margadonna}},
  \bibinfo{author}{\bibfnamefont{K.}~\bibnamefont{Prassides}},
  \bibnamefont{and} \bibinfo{author}{\bibfnamefont{A.~N.} \bibnamefont{Fitch}},
  \bibinfo{journal}{Angewandte Chemie International Edition}
  \textbf{\bibinfo{volume}{43}}, \bibinfo{pages}{6316} (\bibinfo{year}{2004}).

\bibitem[{\citenamefont{Moritomo
  et~al.}(2003{\natexlab{b}})\citenamefont{Moritomo, Kuriki, Ohoyama, Tokoro,
  ichi Ohkoshi, Hashimoto, and Hamada}}]{Moritomo2003}
\bibinfo{author}{\bibfnamefont{Y.}~\bibnamefont{Moritomo}},
  \bibinfo{author}{\bibfnamefont{A.}~\bibnamefont{Kuriki}},
  \bibinfo{author}{\bibfnamefont{K.}~\bibnamefont{Ohoyama}},
  \bibinfo{author}{\bibfnamefont{H.}~\bibnamefont{Tokoro}},
  \bibinfo{author}{\bibfnamefont{S.}~\bibnamefont{ichi Ohkoshi}},
  \bibinfo{author}{\bibfnamefont{K.}~\bibnamefont{Hashimoto}},
  \bibnamefont{and} \bibinfo{author}{\bibfnamefont{N.}~\bibnamefont{Hamada}},
  \bibinfo{journal}{Journal of the Physical Society of Japan}
  \textbf{\bibinfo{volume}{72}}, \bibinfo{pages}{456}
  (\bibinfo{year}{2003}{\natexlab{b}}).

\bibitem[{\citenamefont{Ohkoshi et~al.}(2007)\citenamefont{Ohkoshi, Tokoro,
  Matsuda, Takahashi, Irie, and Hashimoto}}]{Ohkoshi2007}
\bibinfo{author}{\bibfnamefont{S.}~\bibnamefont{Ohkoshi}},
  \bibinfo{author}{\bibfnamefont{H.}~\bibnamefont{Tokoro}},
  \bibinfo{author}{\bibfnamefont{T.}~\bibnamefont{Matsuda}},
  \bibinfo{author}{\bibfnamefont{H.}~\bibnamefont{Takahashi}},
  \bibinfo{author}{\bibfnamefont{H.}~\bibnamefont{Irie}}, \bibnamefont{and}
  \bibinfo{author}{\bibfnamefont{K.}~\bibnamefont{Hashimoto}},
  \bibinfo{journal}{Angewandte Chemie-International Edition}
  \textbf{\bibinfo{volume}{46}}, \bibinfo{pages}{3238} (\bibinfo{year}{2007}).

\bibitem[{\citenamefont{Vertelman et~al.}(2006)\citenamefont{Vertelman,
  Maccallini, Gournis, Rudolf, Bakas, Luzon, Broer, Pugzlys, Lummen, van
  Loosdrecht et~al.}}]{Vertelman2006}
\bibinfo{author}{\bibfnamefont{E.~J.~M.} \bibnamefont{Vertelman}},
  \bibinfo{author}{\bibfnamefont{E.}~\bibnamefont{Maccallini}},
  \bibinfo{author}{\bibfnamefont{D.}~\bibnamefont{Gournis}},
  \bibinfo{author}{\bibfnamefont{P.}~\bibnamefont{Rudolf}},
  \bibinfo{author}{\bibfnamefont{T.}~\bibnamefont{Bakas}},
  \bibinfo{author}{\bibfnamefont{J.}~\bibnamefont{Luzon}},
  \bibinfo{author}{\bibfnamefont{R.}~\bibnamefont{Broer}},
  \bibinfo{author}{\bibfnamefont{A.}~\bibnamefont{Pugzlys}},
  \bibinfo{author}{\bibfnamefont{T.~T.~A.} \bibnamefont{Lummen}},
  \bibinfo{author}{\bibfnamefont{P.~H.~M.} \bibnamefont{van Loosdrecht}},
  \bibnamefont{et~al.}, \bibinfo{journal}{Chem. Mater.}
  \textbf{\bibinfo{volume}{18}}, \bibinfo{pages}{1951} (\bibinfo{year}{2006}).

\bibitem[{\citenamefont{Anderson}(1954)}]{Anderson1954}
\bibinfo{author}{\bibfnamefont{P.~W.} \bibnamefont{Anderson}},
  \bibinfo{journal}{Journal of the Physical Society of Japan}
  \textbf{\bibinfo{volume}{9}}, \bibinfo{pages}{316} (\bibinfo{year}{1954}).

\bibitem[{\citenamefont{Nagy et~al.}(2009)\citenamefont{Nagy, Nafradi, Kushch,
  Yagubskii, Herdtweck, Feher, Kiss, Forro, and Janossy}}]{Nagy2009}
\bibinfo{author}{\bibfnamefont{K.~L.} \bibnamefont{Nagy}},
  \bibinfo{author}{\bibfnamefont{B.}~\bibnamefont{Nafradi}},
  \bibinfo{author}{\bibfnamefont{N.~D.} \bibnamefont{Kushch}},
  \bibinfo{author}{\bibfnamefont{E.~B.} \bibnamefont{Yagubskii}},
  \bibinfo{author}{\bibfnamefont{E.}~\bibnamefont{Herdtweck}},
  \bibinfo{author}{\bibfnamefont{T.}~\bibnamefont{Feher}},
  \bibinfo{author}{\bibfnamefont{L.~F.} \bibnamefont{Kiss}},
  \bibinfo{author}{\bibfnamefont{L.}~\bibnamefont{Forro}}, \bibnamefont{and}
  \bibinfo{author}{\bibfnamefont{A.}~\bibnamefont{Janossy}},
  \bibinfo{journal}{Physical Review B} \textbf{\bibinfo{volume}{80}},
  \bibinfo{pages}{104407} (\bibinfo{year}{2009}).

\end{thebibliography}
\end{document}